\documentclass[12pt]{article}
\usepackage{latexsym}
\usepackage{amsthm}
\usepackage{amssymb}
\usepackage{amsmath}
\usepackage[all]{xy}
\usepackage{graphicx}
\usepackage{subfigure}
\usepackage{dcolumn}
\usepackage{bm}
\usepackage{bbm}
\usepackage[usenames,dvipsnames]{xcolor}
\usepackage{amsfonts}
\usepackage{cases}
\usepackage{float}
\usepackage{lineno}
\usepackage{caption}  
\usepackage{graphicx}
\usepackage{lipsum}   
\usepackage{siunitx}
\usepackage[dvipsnames]{xcolor}

\usepackage{scicite}

\usepackage{amsmath}
\usepackage{amssymb}
\usepackage{mathtools}
\usepackage{times}

\usepackage{graphicx}
\usepackage{caption}
\usepackage{hyperref}
\usepackage{color}
\usepackage{bm}

\newenvironment{sciabstract}{%
\begin{quote} \bf}
{\end{quote}}

\title{Single-cell identification with quantum-enhanced nuclear magnetic resonance}

\author
{Zhiyuan Zhao$^{1,2\dag}$, Qian Shi$^{1,2\dag}$, Shaoyi Xu$^{1}$, Xiangyu Ye$^{1}$, Mengze Shen$^{1}$, Jia Su$^{1}$,\\ Ya Wang$^{1,2,3}$, Tianyu Xie$^{1,2}$, Qingsong Hu$^{5}$, Fazhan Shi$^{1,2,3,4,5\ast}$, Jiangfeng Du$^{3,6\ast}$\\
\\
\normalsize{$^{1}$Laboratory of Spin Magnetic Resonance, School of Physical Sciences,}\\
\normalsize{Anhui Province Key Laboratory of Scientific Instrument Development and Application,}\\
\normalsize{University of Science and Technology of China, Hefei 230026, China}\\
\normalsize{$^{2}$Hefei National Research Center for Physical Sciences at the Microscale,}\\
\normalsize{University of Science and Technology of China, Hefei 230026, China}\\
\normalsize{$^{3}$Hefei National Laboratory, University of Science and Technology of China, Hefei 230088, China}\\
\normalsize{$^{4}$School of Biomedical Engineering and Suzhou Institute for Advanced Research,}\\
	\normalsize{University of Science and Technology of China, Suzhou 215123, China}\\
\normalsize{$^{5}$The First Affiliated Hospital of USTC, Division of Life Sciences and Medicine,}\\
    \normalsize{University of Science and Technology of China, Hefei 230026, China}\\
\normalsize{$^{6}$State Key Laboratory of Ocean Sensing and School of Physics, }\\
\normalsize{Zhejiang University, Hangzhou 310058, China}\\
\\
\normalsize{$^\dag$These authors contributed equally to this work.}\\
\normalsize{$^\ast$E-mail: fzshi@ustc.edu.cn; djf@ustc.edu.cn}
}

\date{}

\begin{document}

\captionsetup[figure]{labelfont={bf},labelsep=period,name={Fig.}}
\baselineskip24pt

\maketitle


\begin{sciabstract}
Identification of individual cells within heterogeneous populations is essential for biomedical research and clinical diagnostics. 
Conventional labeling-based sorting methods, such as fluorescence-activated cell sorting and magnetic-activated cell sorting, enable precise sorting when reliable markers are available. However, their applicability is limited in cells lacking defined markers or sensitive to labeling, as labeling can compromise cellular viability and function.
We present a single-cell identification approach using quantum-enhanced NMR with diamond nitrogen-vacancy centers for label-free detection of intracellular proton ($^1$H) signals.
Using this method, we distinguish two human tumor cell lines by their proton spin-lattice ($T_1$) relaxation times, which serve as a cell-intrinsic physicochemical signature. It lays the groundwork for label-free sorting applications in rare cell analysis, personalized medicine, and single-cell diagnostics.


\end{sciabstract}

\section*{Introduction}
Cells represent fundamental units of biological research, providing insights into physiological functions, disease mechanisms, and therapeutic strategies\cite{rafelski2024establishing,lim2022emerging,regev2017human,haniffa2021roadmap}. Extracting meaningful information from heterogeneous cell populations often requires identifying and isolating specific cell types.\cite{wognum2003identification,abbaszadegan2017isolation,shields2015microfluidic}. Conventional techniques, including fluorescence-activated cell sorting (FACS)\cite{bonner1972fluorescence,herzenberg2002history} and single-cell sequencing technologies\cite{baysoy2023technological}, have significantly advanced single-cell studies.  However, while FACS depends on known cell-surface markers and often requires antibody-based fluorescent labeling, single-cell sequencing approaches, though not necessarily label-dependent, still rely on pre-isolated cell populations and cannot preserve native cellular states\cite{frenea2022basic}. Consequently, cells lacking distinctive markers or those whose physiological conditions may be perturbed during processing remain challenging to identify and isolate accurately\cite{faraghat2017high}. These limitations underscore the need for developing label-free, live-cell single-cell characterization technologies.

Nuclear magnetic resonance (NMR), a technique that detects various nuclear spins in biological samples, offers label-free, non-destructive access to molecular composition and dynamics in living cells\cite{inomata2009high,barbieri2016characterization,palmer2004nmr}. These features have underpinned its widespread use in structural biology and clinical diagnosis. 
However, conventional NMR lacks the spatial resolution required for single-cell analysis. 
Most measurements rely on ensemble averaging across large populations, obscuring cellular heterogeneity\cite{glover2002limits,luchinat2017cell,theillet2022cell}.
Cell-resolved measurements have been demonstrated only for giant cells at the nL scale ($\sim$100 \textmu m), whereas typical mammalian somatic cells at the pL scale ($\sim$10 \textmu m) remain beyond reach\cite{sivelli2025micro}.

\begin{figure}
\centering
\includegraphics[width=1.0\columnwidth]{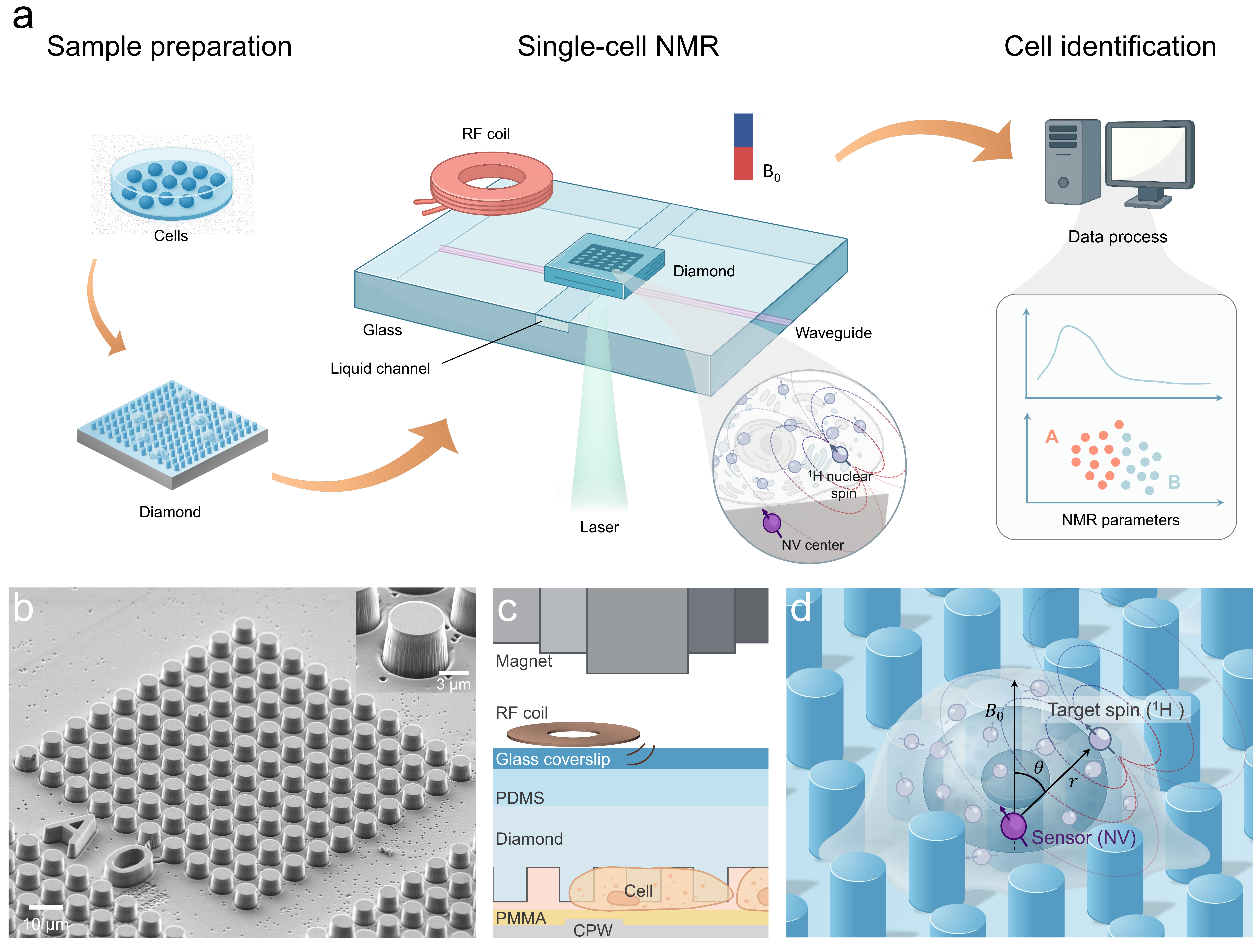}
    \caption{\textbf{NV-enhanced single-cell NMR for label-free cell identification.}
(a) Single-cell identification workflow. Individual cells are measured with NV-enhanced NMR, where NMR contrasts such as proton spin-lattice relaxation enable label-free discrimination of cell types.
A diamond-based microfluidic chip with a liquid channel is aligned below a diamond hosting near-surface NV centers and a microstructured surface region. NV spins are optically initialized and read out with laser pulses. Microwaves, delivered via a coplanar waveguide, drive the NV electron spin, while a radio-frequency (RF) coil addresses intracellular $^{1}\mathrm{H}$ nuclei . A static magnetic field $B_0$ sets the quantization axis. 
Time-domain signals are processed to extract quantitative NMR parameters, which serve as features for label-free classification.
(b) Scanning electron micrograph of the diamond pillar array. Inset, higher-magnification view of a single diamond pillar.
(c) Cross-sectional schematic of the single-cell NMR chip, showing the magnet, RF coil, glass coverslip, polydimethylsiloxane (PDMS) layer, diamond with pillar array, polymethyl methacrylate (PMMA) support and coplanar waveguide (CPW). Further structural details are provided in the Supplementary Information.
(d) Principle of NV-based single-cell NMR, where an NV sensor in a diamond pillar detects the nuclear spins ($^1$H) of a single cell under an external magnetic field. Enlarged view of the dipolar interaction between an NV spin and nuclear spins within a cell. The interaction depends on the NV-nucleus separation $r$ and the relative angle $\theta$ to magnetic field (see Supplementary Information).
 }
    \label{theory}
\end{figure}

Recent advancements in quantum technologies, particularly nitrogen-vacancy (NV) centers in diamond, have transformed magnetic sensing by dramatically improving both sensitivity and spatial resolution\cite{wolf2015subpicotesla,fang2013high,arai2015fourier,zhao2023sub}.  
NV-based quantum sensors enable nanoscale detection\cite{du2024single,mamin2013nanoscale,staudacher2013nuclear,lovchinsky2016nuclear,shi2015single} and chemical resolved spectra\cite{aslam2017nanoscale,schmitt2017submillihertz,glenn2018high}, offering an unparalleled platform for cellular-scale analysis. 
Despite this progress, the application of such technologies to single-cell analysis has remained elusive.
Here, we demonstrate NV-based single-cell NMR, enabling a label-free strategy for cell identification (Fig.~\ref{theory}). We validate the strategy in HeLa and MCF7 cells by measuring proton spin-lattice $T_1$ relaxation as the first example.
This approach links intrinsic physicochemical signatures to cell identity and opens new opportunities in fundamental biological research and clinical diagnostics, including rare cell analysis\cite{proserpio2016single,chen2014rare}, personalized medicine\cite{beckman2012impact,dutta2022single}, and single-cell diagnostics\cite{jovic2022single,potter2018single}.
\subsection*{Principles of single-cell nuclear magnetic resonance}
In biological systems, the dominant elements carbon, hydrogen, nitrogen and phosphorus ($^{13}$C, $^1$H, $^{14}$N, and $^{31}$P) naturally possess nuclear spins. 
Among them, proton ($^1$H) is the most abundant nuclear spin, arising from intracellular water molecules and all biomolecules. 
Single-cell nuclear magnetic resonance (NMR) exploits the magnetic dipolar interaction between these nuclear spins and the electronic spin of a diamond NV sensor. 
Under an external magnetic field aligned with the NV axis, the NV sensor-target nuclear coupling Hamiltonian for a single nucleus is given by
\begin{equation}
\begin{aligned}
H = & \underbrace{DS_z^2 + \gamma_e B_0 S_z \vphantom{\sum^N}}_{\text{Sensor}}
+ \underbrace{S_z  \left(a_{\parallel} I_{z} + a_{\perp} I_{\perp}\right) \vphantom{\sum^N}}_{\text{Sensor-target coupling}}
+ \underbrace{\gamma_n B_0  I_{z} \vphantom{\sum^N}}_{\text{Target nuclear spin}},
\end{aligned}
\end{equation}
where $D$ is the zero-field splitting, $\gamma_e$ and $\gamma_n$ are the electronic and nuclear gyromagnetic ratios, $S_z$ denotes the NV electron spin operator, and $I_z$, $I_{\perp}$ are the spin components with longitudinal and transverse dipolar couplings $a_{\parallel}$ and $a_{\perp}$.
Here $I_{\perp}=\cos{\phi}I_{x}+\sin{\phi}I_{y}$, where $\phi$ and $\theta$ denote the azimuthal and polar angles between the NV axis and the nuclear spin. Neglecting Fermi contact terms, the longitudinal dipolar coupling strength is given by
\begin{equation}
\begin{gathered}
\begin{aligned}
a_{\|}&=\frac{\mu_0 \gamma_{\mathrm{e}} \gamma_{\mathrm{n}} \hbar}{4 \pi r^3}(3 \cos ^2 \theta-1), \\[6pt]
\end{aligned}
\end{gathered}
\end{equation}
where $r$ is the NV-nucleus distance (see Fig.~\ref{theory}d).
The longitudinal coupling can be interpreted as an effective static magnetic field along the NV axis\cite{du2024single}. 
For multiple nuclei, the total Hamiltonian is obtained by summing over all nuclear spins, as detailed in the Supplementary Information. 

\subsection*{Experimental implementation of single-cell NMR}
\begin{figure}
\centering
\includegraphics[width=0.8\columnwidth]{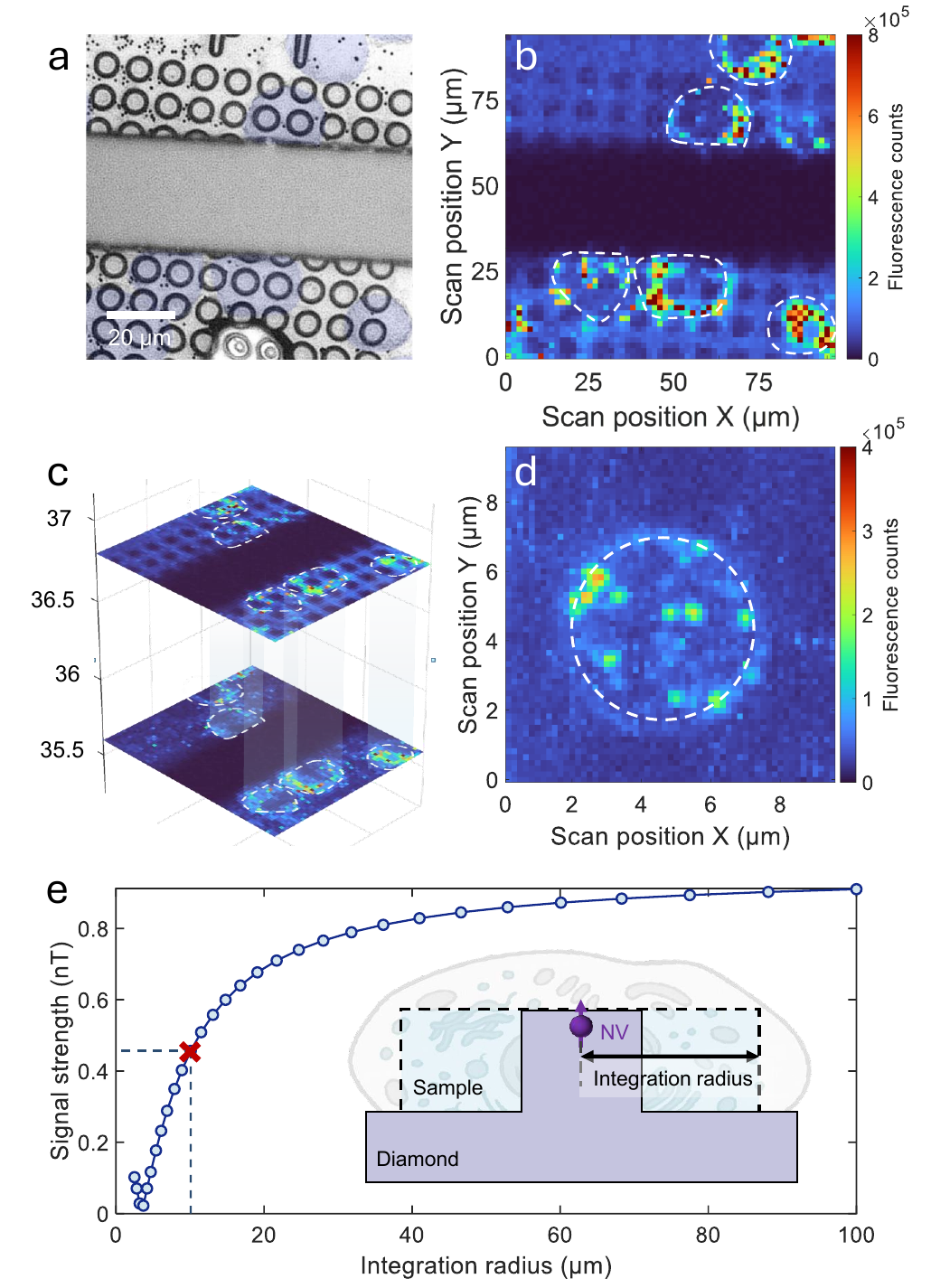}
  \captionof{figure}{\textbf{Localization of the detected NMR signal to a single cell.}
(a) Optical micrograph of the diamond pillar array with cells sealed on top. Cells are highlighted with a blue pseudo-color overlay and the central band corresponds to the CPW.
(b) NV fluorescence scan of the same region as a function of the lateral scan position, with dashed contours indicating the positions of individual cells.
(c) Three-dimensional stack of NV fluorescence images at different focal planes, showing that the fluorescence contrast is confined to the NV pillars located beneath a single cell.
(d) Zoom-in of the NV fluorescence map around one selected pillar used for single-cell measurements.
(e) Integrated NMR signal strength as a function of the lateral integration radius around the NV sensor. The red marker indicates the radius at which half of the total signal is accumulated and the signal saturates for radii comparable to the cell size, demonstrating that the detected NMR signal is dominated by nuclear spins within a single cell. }
\label{cell}
\end{figure}

Single-cell NMR is implemented on diamond micropillars that host individual near-surface NV centres at their tips, thereby providing intrinsically localised sensing volumes and avoiding the spatial averaging inherent in NV ensembles.
A thin PDMS layer applies mild mechanical pressure that gently flattens cells onto the pillar array, improving contact with the diamond surface while expanding the lateral footprint of individual cells. The relative positions of cells and pillars are registered by optical microscopy and verified by confocal NV fluorescence scans at different focal planes, confirming that the pillars selected for NMR detectiont are fully enveloped by a single cell (Fig. \ref{cell}a–d). 
To quantify the spatial origin of the detected signal, we evaluate the NV-detected NMR signal as a function of the lateral integration radius around a given pillar. The integrated signal saturates at a radius comparable to the projected cell size, indicating that the dominant contribution arises from nuclei within a single cell (Fig. \ref{cell}e).

While the NV-proton coupling provides the physical basis for detection, practical realization of single-cell NMR demands strategies to overcome inherently weak signals. Accordingly, enhancing the signal-to-noise ratio (SNR) requires both improvement of magnetic sensitivity and amplification of the nuclear signal. To enhance sensitivity while preserving spatial resolution, we used single NV centers in isotopically purified $^{12}$C diamond\cite{han2025solid} which provides millisecond-scale coherence times, together with real-time feedback charge-state preparation\cite{hopper2020real,xie2021beating} and repetitive readout protocols\cite{neumann2010single} to improve initialization and readout fidelity. Overhauser dynamic nuclear polarization (DNP)\cite{carver1956experimental,bucher2020hyperpolarization} increased nuclear spin polarization thereby amplifying the nuclear signal. 

\begin{figure}
\centering
\includegraphics[width=0.9\columnwidth]{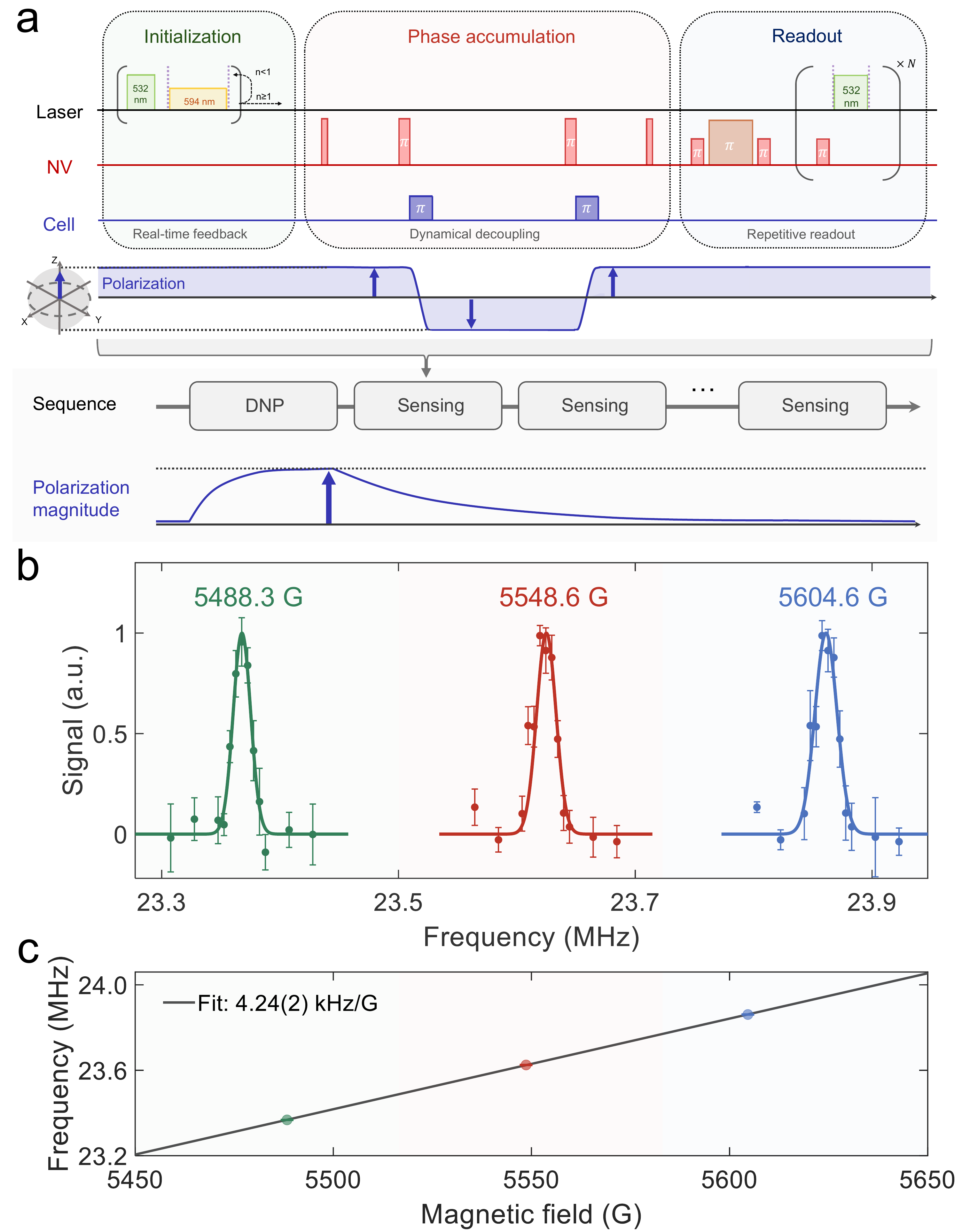}
  \captionof{figure}{\textbf{Single-cell NMR detection and identification of proton signals.}
(a) Top: The detection sequence consists of three parts: initialization with real-time feedback charge-state preparation, the ENDOR sequence with dynamical decoupling, and readout via SWAP operations that transfer the electron state to the nuclear spin for repetitive readout. Bottom: The full experiment begins with DNP to build nuclear polarization, followed by repeated sensing cycles. Polarization is sampled in each cycle and relaxes between cycles toward Boltzmann equilibrium with $T_1$.
(b) Proton resonance spectra obtained at different magnetic fields, with the horizontal axis showing the applied RF frequency. Solid lines are fits to the data. 
(c) Linear fit of resonance frequency versus magnetic field yields a slope of 4.24(2) kHz/G, in agreement with the proton gyromagnetic ratio (4.26 kHz/G).
}
\label{experiment1}
\end{figure}

Analogous to an optical interferometer, we employ a spin-based quantum interferometer to detect the proton signal. The nuclear magnetic signal detection sequence (Fig.~\ref{experiment1}a) consists of three stages: initialization, phase accumulation, and readout. 
In the initialization stage, real-time feedback charge-state initialization is employed to simultaneously prepare both the charge and spin states of the NV center.
The second stage employs an electron-nuclear double resonance (ENDOR) sequence: a $\pi_x/2$ pulse prepares the NV electron spin in the $(|0\rangle+|1\rangle)/2$ state, followed by two $\pi$ pulses applied to the electron spin while subsequent radio-frequency $\pi$ pulses are applied to the sample nuclei. 
In this way, low-frequency environmental noise is suppressed while the NV-nuclear spin coupling is preserved. 
To suppress phase accumulation induced by radio-frequency (RF) control, a two-$\pi$ pulse echo sequence is employed.
Finally, a $\pi_y/2$ pulse is applied to read out the polarization signal,
\begin{equation}
P_{\mathrm{pola}} \approx \frac{1}{2}+\frac{P\sum_i a_{\|, i} \tau}{4},
\end{equation}
where $P$ is the nuclear spin polarization. Because the polarization generated by DNP is opposite to the thermal equilibrium polarization, continuous measurements following DNP yield the spin-lattice relaxation curve of protons. To prevent depolarization induced by imperfect control, adiabatic RF pulses were applied, and the amplitude was precisely calibrated. Calibration was performed by measuring the RF-induced Bloch-Siegert shift of the NV energy levels.
In the final readout stage, the electronic state is transferred to the nuclear spin, enabling repetitive readout of the NV spin state. Further details of the experimental sequence are provided in the Supplementary Information.


Sweeping the RF frequency in the sequence yields the nuclear magnetic resonance spectrum in Fig.~\ref{experiment1}b. To confirm the signal originates from protons, we performed field-dependent measurements. By tracking the resonance positions at three magnetic field strengths, we obtained a slope of 4.24(2) kHz/G, in excellent agreement with the proton gyromagnetic ratio of 4.26 kHz/G. This confirms that the detected nuclear magnetic signal arises from intracellular proton spins.  The linewidth is narrower than the power-broadened width of a single RF pulse because correlations between repetitions allow off-resonant pulse errors to accumulate, narrowing the line.

\subsection*{Validation of cell identification by single-cell NMR}
Early NMR studies revealed tissue-dependent differences in nuclear spin relaxation\cite{damadian1971tumor}, providing the physical and diagnostic basis for magnetic resonance imaging (MRI). Building on similar principle, we present a label-free strategy for cell identification using single-cell NMR contrasts and validate it by discriminating cell types via proton spin-lattice ($T_1$) relaxation measured in individual cells.

\begin{figure}
\centering
\includegraphics[width=0.99\columnwidth]{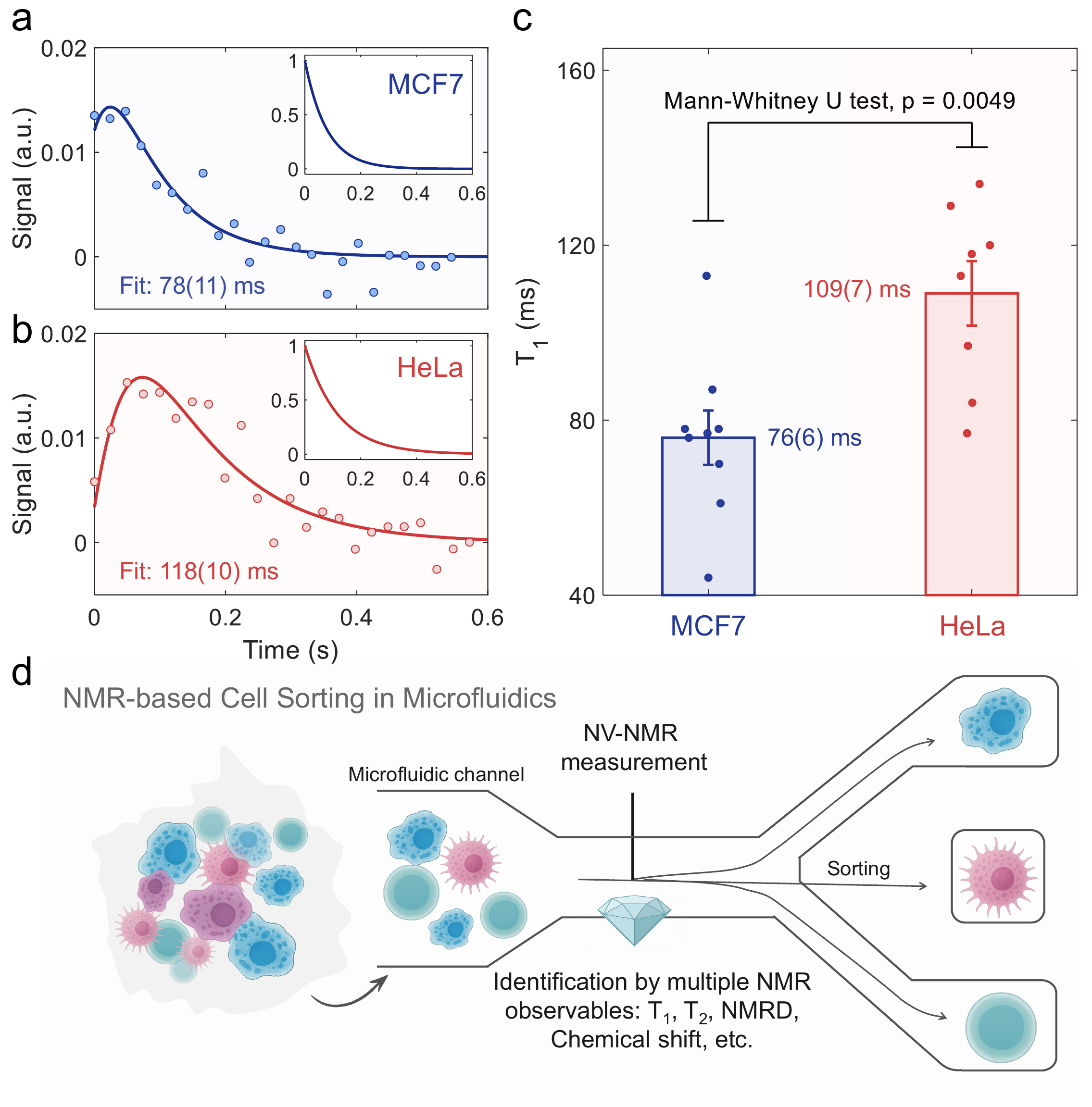}
    \caption{\textbf{Measurement and statistical analysis of single-cell relaxation times for label-free cell identification.}
(a) Representative $T_1$ relaxation curves for an individual MCF7 cell. Markers show a three-point moving average of the raw signal. Because the detected NMR signal exceeds the linear regime, solid curves are fits to the function $C_0\sin(Ae^{-t/T_1})$. The inset shows the reconstructed polarization decay $P(t)\propto e^{-t/T_1}$.
(b) Representative $T_1$ relaxation curves for an individual HeLa cell. 
(c) Statistical comparison of nuclear relaxation times. Population statistics of $T_1$ show a significant difference between MCF7 and HeLa cells ($p=0.0049$, Mann-Whitney U test), with a large effect size (Cliff's $\delta=0.78$) and a median $T_1$ difference of about 36.5 ms (Hodges-Lehmann estimator). Mean $T_1$ values are $76(6)$ ms for MCF7 and $109(7)$ ms for HeLa. 
(d) Envisioned multi-parametric single-cell analysis with NMR-based microfluidics. Cells flow through microfluidic channels into an NV-based detection region, where intrinsic NMR signatures ($T_1$, $T_2$, chemical shift, NMRD) are measured for label-free identification and sorting. This platform can be extended to high-throughput analysis and integration with other single-cell omics.
 }
\label{experiment2}
\end{figure}

MCF7 cells\cite{comcsa2015story}, derived from human breast adenocarcinoma, are widely used as a model in cancer biology, whereas HeLa cells\cite{masters2002hela}, originating from cervical carcinoma, are among the most extensively studied human cell lines. These two cell types were selected for single-cell $T_1$ measurements to assess whether relaxation times can distinguish them.
Fig. \ref{experiment2}a and b show representative single-cell proton spin-lattice decays from an MCF7 cell and a HeLa cell, respectively, illustrating their different relaxation rates.
The average $T_1$ for protons in MCF7 cells was $76(6)$ ms, whereas HeLa cells exhibited a longer average $T_1$ of $109(7)$ ms (Fig. \ref{experiment2}c).
Single-cell $T_1$ relaxation distinguishes MCF7 and HeLa populations ($p=0.0049$), with a large effect size ($\delta=0.78$) and a median $T_1$ difference of 36.5 ms.
The $T_1$ distributions differ significantly between the two cell populations, demonstrating that single-cell $T_1$ relaxation times can be used to discriminate between distinct cell types.

\section*{Conclusion and outlook}
We demonstrate label-free cell identification with NV-enhanced single-cell NMR.
Measuring intracellular proton spin-lattice relaxation enables reliable discrimination between cell types without exogenous labels, thereby eliminating the dependence on cell-surface markers. These results establish $T_1$ as an intrinsic physicochemical signature of the intracellular microenvironment and position single-cell NMR as a physics-based modality for label-free classification.

At the cellular level, the spin-lattice relaxation time of intracellular water provides an integrated measure of the physicochemical microenvironment. Variations in $T_1$ arise from the interplay of protein-water interfaces, hydration dynamics, paramagnetic radicals and the ionic milieu\cite{fullerton1984frequency,bottomley1984review,bottomley1987review,koenig1996molecular,gaeta2021magnetism}. 
Analysis of our single-cell NMR data demonstrates that the spin-lattice relaxation time can discriminate between different cell types. 
To enhance discriminative power across diverse cell types and in highly heterogeneous samples, a natural extension is to incorporate additional NMR dimensions which are sensitive to complementary physicochemical variables, such as spin-spin relaxation time ($T_2$), chemical shifts, nuclear magnetic relaxation dispersion (NMRD)\cite{koenig1990field}, magnetization transfer\cite{henkelman2001magnetization}.
These multi-parametric NMR contrasts provide complementary sensitivity to susceptibility-induced dephasing, molecular composition, and rotational dynamics \cite{nitz1999contrast}.

Our array-based implementation affords detailed single-cell measurements, but in its current form it only permits buffer solution to flow across the pillars rather than transporting cells, resulting in limited throughput and constraining population-scale analyses of heterogeneity. This limitation is imposed by the present fluidic design rather than by the NV-based detection itself. Integrating single-cell NMR with more advanced microfluidic platforms\cite{whitesides2006origins,squires2005microfluidics,allert2022microfluidic} offers a promising route to truly high-throughput, massively parallel measurements and large-scale datasets, paving the way for a single-cell NMR atlas (Fig. \ref{experiment2}d).


Single-cell NMR provides unique access to relaxation, spectroscopic, and spatial information from the cell interior. With single-NV implementations, it can approach subcellular resolution and integrate with single-cell sequencing, single-cell transcriptomics, single-cell proteomics and single-cell metabolomics for multimodal cellular characterization\cite{baysoy2023technological}. Such an approach could reshape the definition and classification of cell states, providing a physics-based complement to genomics and proteomics with implications from fundamental cell biology to precision medicine.


\section*{Methods}
\subsection*{Diamond sensor preparation}
\label{experimental_setup}
The diamond substrate used in this work was a \SI{50}{\micro\meter}-thick, (111)-oriented diamond of natural isotopic abundance.
Its surface layer was grown by chemical vapor deposition and isotopically purified to 99.99\% $^{12} $C to suppress spin-bath noise. NV centers were created by 60-keV $^{15}$N$^+$ ion implantation followed by annealing at 1000 $^{\circ}\mathrm{C}$. An array of etched micropillars (5-\textmu m height, 5-\textmu m diameter, 10-\textmu m pitch) was fabricated on the diamond surface.

\subsection*{Experimental setup}
The simplified experimental setup, illustrated in Fig. S1, can be divided into four main sections. The first part is microwave and radio-frequency system. Microwave control pulses for the NV centers were generated using an arbitrary waveform generator (AWG, Keysight M8190) and a PSG Vector Signal Generator (Keysight E8267D). The MW frequency was up-converted through an IQ modulator before being applied to NV center. For Overhauser DNP, a signal generator (HMC-T2220) delivered microwave excitation at the electron spin resonance frequency of TEMPOL.
For RF control, two additional AWGs (Keysight 33522B and 33622A) generated independent pulse sequences. One addressed the NV's native nitrogen nuclear spin. Its output was combined with the microwave drive via a diplexer and delivered to the coplanar waveguide. The other drove a compact RF coil for $^1$H manipulation through a home-built RLC circuit. All microwave and RF paths were amplified by separate power amplifiers before being applied to the sample.
Laser excitation used 532- and 594-nm sources. Both beams were gated with acousto-optic modulators, combined using an RGB combiner, and directed by a dichroic beam splitter into an oil-immersion objective (Olympus, UPLAPO100XOHR) to focus onto the diamond. Photoluminescence from the NV centers was collected through the same objective, spectrally separated by the same dichroic, and detected with an avalanche photodiode.
The real-time feedback and control unit consisted of a Zurich Instruments AWG and a PulseBlaster timing controller. During charge-state pre-selection, the AWG ran a photon-count-based feedback loop in which photon counts served as decision signals to proceed or restart, ensuring initialization to NV$^-$. The PulseBlaster synchronized all AWG channels, AOMs, and hardware triggers, controlling the timing of the entire experimental sequence.
As shown in Fig. S1, the sample, coil, objective, and a single-sided magnet mounted on a translation stage were housed in a temperature-controlled enclosure to ensure magnetic-field and temperature stability during the experiments. The field homogeneity of the single-sided Halbach magnet is shown in Fig. S5.

\subsection*{Cell preparation and mounting}
MCF7 (ATCC HTB-22) and HeLa (ATCC CCL-2) cell lines were purchased from Pricella, authenticated by short tandem repeat (STR) analysis, and tested negative for mycoplasma. Cells were cultured in MEM (Pricella) supplemented with 10\% fetal bovine serum (Vivacell) and 1\% penicillin-streptomycin (HyClone) at \SI{37}{\celsius} in a humidified incubator with 5\% CO$_2$. The MCF7 cultures were additionally supplemented with \SI{10}{\micro\gram\per\milli\liter} insulin.
The cells were washed with phosphate-buffered saline (PBS), and then dissociated with trypsin (Thermo Fisher Scientific) to obtain a cell suspension. Subsequently, 100 mM TEMPOL in 1× HEPES buffer was added to the cell suspension at a 1:4 (v/v) ratio (buffer:cell suspension). The cell suspension was then dropped onto the waveguide structure and covered with the diamond. Afterwards, the assembly was rinsed three times with 20 mM TEMPOL in 1× HEPES buffer and sealed with wax. Finally the sample was mounted on the platform for measurement. A simplified schematic of the workflow is shown in Fig. S2.

\subsection*{Data processing}
The raw fluorescence was acquired in two detection phases by alternating the final ENDOR readout pulse between $+\pi_y/2$ and $-\pi_y/2$, corresponding to $N_{+}$ and $N_{-}$.
For each measurement, the signal contrast was defined as
\begin{equation}
  C=\frac{N_{+}-N_{-}}{N_{+}},
\end{equation}
which compensates for fluctuations in fluorescence. Each experiment comprised 70 repeats following a DNP sequence to abtain the $T_1$ relaxation data, and the entire measurement was repeated multiple times to improve the signal-to-noise ratio. For each individual cell, the experiment was performed over detuning values in the range of [-60,60] kHz. 
The central region [-15,15] kHz was taken as the signal window, while the outer detuning points served as references. The baseline-corrected contrast was computed as
\begin{equation}
  C_{\mathrm{corr}}(\tau) =  \overline{C}_{\mathrm{win}}(\tau) - \overline{C}_{\mathrm{ref}}(\tau),
\end{equation}
where $\overline{C}_{\mathrm{win}}(\tau)$ is the average over the window and $\overline{C}_{\mathrm{ref}}(\tau)$ is the average over the outer detuning points.
The $T_1$ values were extracted by fitting the data to a single-exponential decay function:
\begin{equation}
  C(\tau)=C_0 \sin(A*e^{-\tau/T_1})
\end{equation}
rather than a simple exponential decay, which reflects the detected NMR signal exceeds the linear regime.
When the signal is sufficiently strong, the accumulated phase can exceed $\pi/2$, leading to an apparent inversion of the measured contrast.

\subsection*{Statistical analysis}
The $T_1$ of MCF7 and Hela cells ($T_1^\text{M}$ and $T_1^\text{H}$) were compared using the exact two-sided Mann-Whitney U test ($p=0.0049$).
Effect size was summarized with Cliff's delta ($\delta=0.78$, 95\% CI 0.40-1.00), with confidence limits obtained via nonparametric bootstrap resampling.
We also reported the Hodges-Lehmann estimator of the median difference defined as ($T_1^\text{M}-T_1^\text{H}$) (estimate -36.5, 95\% CI -52.5 to -8.5).

\bibliographystyle{ScienceAdvances}
\bibliography{references}

@article{rafelski2024establishing,
  title={Establishing a conceptual framework for holistic cell states and state transitions},
  author={Rafelski, Susanne M and Theriot, Julie A},
  journal={Cell},
  volume={187},
  number={11},
  pages={2633--2651},
  year={2024},
  publisher={Elsevier}
}

@article{lim2022emerging,
  title={The emerging era of cell engineering: {Harnessing} the modularity of cells to program complex biological function},
  author={Lim, Wendell A},
  journal={Science},
  volume={378},
  number={6622},
  pages={848--852},
  year={2022},
  publisher={American Association for the Advancement of Science}
}

@article{regev2017human,
  title={The human cell atlas},
  author={Regev, Aviv and Teichmann, Sarah A and Lander, Eric S and Amit, Ido and Benoist, Christophe and Birney, Ewan and Bodenmiller, Bernd and Campbell, Peter and Carninci, Piero and Clatworthy, Menna and others},
  journal={elife},
  volume={6},
  pages={e27041},
  year={2017},
  publisher={eLife Sciences Publications, Ltd}
}

@article{haniffa2021roadmap,
  title={A roadmap for the human developmental cell atlas},
  author={Haniffa, Muzlifah and Taylor, Deanne and Linnarsson, Sten and Aronow, Bruce J and Bader, Gary D and Barker, Roger A and Camara, Pablo G and Camp, J Gray and Ch{\'e}dotal, Alain and Copp, Andrew and others},
  journal={Nature},
  volume={597},
  number={7875},
  pages={196--205},
  year={2021},
  publisher={Nature Publishing Group UK London}
}

@article{wognum2003identification,
  title={Identification and isolation of hematopoietic stem cells},
  author={Wognum, Albertus W and Eaves, Allen C and Thomas, Terry E},
  journal={Archives of medical research},
  volume={34},
  number={6},
  pages={461--475},
  year={2003},
  publisher={Elsevier}
}

@article{abbaszadegan2017isolation,
  title={Isolation, identification, and characterization of cancer stem cells: {A} review},
  author={Abbaszadegan, Mohammad Reza and Bagheri, Vahid and Razavi, Mahya Shariat and Momtazi, Amir Abbas and Sahebkar, Amirhossein and Gholamin, Mehran},
  journal={Journal of cellular physiology},
  volume={232},
  number={8},
  pages={2008--2018},
  year={2017},
  publisher={Wiley Online Library}
}

@article{shields2015microfluidic,
  title={Microfluidic cell sorting: a review of the advances in the separation of cells from debulking to rare cell isolation},
  author={Shields Iv, C Wyatt and Reyes, Catherine D and L{\'o}pez, Gabriel P},
  journal={Lab on a Chip},
  volume={15},
  number={5},
  pages={1230--1249},
  year={2015},
  publisher={Royal Society of Chemistry}
}

@article{herzenberg2002history,
  title={The history and future of the fluorescence activated cell sorter and flow cytometry: a view from {Stanford}},
  author={Herzenberg, Leonard A and Parks, David and Sahaf, Bita and Perez, Omar and Roederer, Mario and Herzenberg, Leonore A},
  journal={Clinical chemistry},
  volume={48},
  number={10},
  pages={1819--1827},
  year={2002},
  publisher={Oxford University Press}
}

@article{bonner1972fluorescence,
  title={Fluorescence activated cell sorting},
  author={Bonner, WA and Hulett, HR and Sweet, RG and Herzenberg, LA},
  journal={Review of Scientific Instruments},
  volume={43},
  number={3},
  pages={404--409},
  year={1972},
  publisher={American Institute of Physics}
}

@article{frenea2022basic,
  title={Basic principles and recent advances in magnetic cell separation},
  author={Frenea-Robin, Marie and Marchalot, Julien},
  journal={Magnetochemistry},
  volume={8},
  number={1},
  pages={11},
  year={2022},
  publisher={MDPI}
}

@article{faraghat2017high,
  title={High-throughput, low-loss, low-cost, and label-free cell separation using electrophysiology-activated cell enrichment},
  author={Faraghat, Shabnam A and Hoettges, Kai F and Steinbach, Max K and Van Der Veen, Daan R and Brackenbury, William J and Henslee, Erin A and Labeed, Fatima H and Hughes, Michael P},
  journal={Proceedings of the National Academy of Sciences},
  volume={114},
  number={18},
  pages={4591--4596},
  year={2017},
  publisher={National Academy of Sciences}
}

@article{fullerton1984frequency,
  title={Frequency dependence of magnetic resonance spin-lattice relaxation of protons in biological materials.},
  author={Fullerton, GD and Cameron, IL and Ord, VA},
  journal={Radiology},
  volume={151},
  number={1},
  pages={135--138},
  year={1984}
}

@article{bottomley1984review,
  title={A review of normal tissue hydrogen {NMR} relaxation times and relaxation mechanisms from 1--100 {MHz}: dependence on tissue type, {NMR} frequency, temperature, species, excision, and age},
  author={Bottomley, Paul A and Foster, Thomas H and Argersinger, Raymond E and Pfeifer, Leah M},
  journal={Medical physics},
  volume={11},
  number={4},
  pages={425--448},
  year={1984},
  publisher={Wiley Online Library}
}

@article{bottomley1987review,
  title={A review of {$^1$H} nuclear magnetic resonance relaxation in pathology: are {T$_1$} and {T$_2$} diagnostic?},
  author={Bottomley, Paul A and Hardy, CJ and Argersinger, RE and Allen-Moore, G},
  journal={Medical physics},
  volume={14},
  number={1},
  pages={1--37},
  year={1987},
  publisher={Wiley Online Library}
}

@article{koenig1996molecular,
  title={Molecular basis of magnetic relaxation of water protons of tissue},
  author={Koenig, Seymour H},
  journal={Academic radiology},
  volume={3},
  number={7},
  pages={597--606},
  year={1996},
  publisher={Elsevier}
}

@article{gaeta2021magnetism,
  title={Magnetism of materials: theory and practice in magnetic resonance imaging},
  author={Gaeta, Michele and Cavallaro, Marco and Vinci, Sergio Lucio and Mormina, Enricomaria and Blandino, Alfredo and Marino, Maria Adele and Granata, Francesca and Tessitore, Agostino and Galletta, Karol and D’Angelo, Tommaso and others},
  journal={Insights into Imaging},
  volume={12},
  number={1},
  pages={179},
  year={2021},
  publisher={Springer}
}

@article{glover2002limits,
  title={Limits to magnetic resonance microscopy},
  author={Glover, Paul and Mansfield, Peter},
  journal={Reports on progress in physics},
  volume={65},
  number={10},
  pages={1489},
  year={2002},
  publisher={IOP Publishing}
}

@article{luchinat2017cell,
  title={In-cell {NMR}: a topical review},
  author={Luchinat, Enrico and Banci, Lucia},
  journal={IUCrJ},
  volume={4},
  number={2},
  pages={108--118},
  year={2017},
  publisher={International Union of Crystallography}
}

@article{theillet2022cell,
  title={In-cell {NMR}: {Why} and how?},
  author={Theillet, Francois-Xavier and Luchinat, Enrico},
  journal={Progress in Nuclear Magnetic Resonance Spectroscopy},
  volume={132},
  pages={1--112},
  year={2022},
  publisher={Elsevier}
}

@article{palmer2004nmr,
  title={{NMR} characterization of the dynamics of biomacromolecules},
  author={Palmer III, Arthur G},
  journal={Chemical reviews},
  volume={104},
  number={8},
  pages={3623--3640},
  year={2004},
  publisher={ACS Publications}
}

@article{inomata2009high,
  title={High-resolution multi-dimensional {NMR} spectroscopy of proteins in human cells},
  author={Inomata, Kohsuke and Ohno, Ayako and Tochio, Hidehito and Isogai, Shin and Tenno, Takeshi and Nakase, Ikuhiko and Takeuchi, Toshihide and Futaki, Shiroh and Ito, Yutaka and Hiroaki, Hidekazu and others},
  journal={Nature},
  volume={458},
  number={7234},
  pages={106--109},
  year={2009},
  publisher={Nature Publishing Group UK London}
}

@article{barbieri2016characterization,
  title={Characterization of proteins by in-cell {NMR} spectroscopy in cultured mammalian cells},
  author={Barbieri, Letizia and Luchinat, Enrico and Banci, Lucia},
  journal={Nature protocols},
  volume={11},
  number={6},
  pages={1101--1111},
  year={2016},
  publisher={Nature Publishing Group UK London}
}

@article{wolf2015subpicotesla,
  title={Subpicotesla diamond magnetometry},
  author={Wolf, Thomas and Neumann, Philipp and Nakamura, Kazuo and Sumiya, Hitoshi and Ohshima, Takeshi and Isoya, Junichi and Wrachtrup, J{\"o}rg},
  journal={Physical Review X},
  volume={5},
  number={4},
  pages={041001},
  year={2015},
  publisher={APS}
}

@article{fang2013high,
  title={High-sensitivity magnetometry based on quantum beats in diamond nitrogen-vacancy centers},
  author={Fang, Kejie and Acosta, Victor M and Santori, Charles and Huang, Zhihong and Itoh, Kohei M and Watanabe, Hideyuki and Shikata, Shinichi and Beausoleil, Raymond G},
  journal={Physical review letters},
  volume={110},
  number={13},
  pages={130802},
  year={2013},
  publisher={APS}
}

@article{lovchinsky2016nuclear,
  title={Nuclear magnetic resonance detection and spectroscopy of single proteins using quantum logic},
  author={Lovchinsky, Igor and Sushkov, AO and Urbach, Elana and de Leon, Nathalie P and Choi, Soonwon and De Greve, Kristiaan and Evans, Ruffin and Gertner, Rona and Bersin, Eric and M{\"u}ller, C and others},
  journal={Science},
  volume={351},
  number={6275},
  pages={836--841},
  year={2016},
  publisher={American Association for the Advancement of Science}
}

@article{zhao2023sub,
  title={Sub-nanotesla sensitivity at the nanoscale with a single spin},
  author={Zhao, Zhiyuan and Ye, Xiangyu and Xu, Shaoyi and Yu, Pei and Yang, Zhiping and Kong, Xi and Wang, Ya and Xie, Tianyu and Shi, Fazhan and Du, Jiangfeng},
  journal={National Science Review},
  volume={10},
  number={12},
  pages={nwad100},
  year={2023},
  publisher={Oxford University Press}
}

@article{arai2015fourier,
  title={Fourier magnetic imaging with nanoscale resolution and compressed sensing speed-up using electronic spins in diamond},
  author={Arai, Keigo and Belthangady, Chinmay and Zhang, Huiliang and Bar-Gill, N and DeVience, SJ and Cappellaro, Paola and Yacoby, Amir and Walsworth, Ronald Lee},
  journal={Nature nanotechnology},
  volume={10},
  number={10},
  pages={859--864},
  year={2015},
  publisher={Nature Publishing Group UK London}
}

@article{shi2015single,
  title={Single-protein spin resonance spectroscopy under ambient conditions},
  author={Shi, Fazhan and Zhang, Qi and Wang, Pengfei and Sun, Hongbin and Wang, Jiarong and Rong, Xing and Chen, Ming and Ju, Chenyong and Reinhard, Friedemann and Chen, Hongwei and others},
  journal={Science},
  volume={347},
  number={6226},
  pages={1135--1138},
  year={2015},
  publisher={American Association for the Advancement of Science}
}

@article{glenn2018high,
  title={High-resolution magnetic resonance spectroscopy using a solid-state spin sensor},
  author={Glenn, David R and Bucher, Dominik B and Lee, Junghyun and Lukin, Mikhail D and Park, Hongkun and Walsworth, Ronald L},
  journal={Nature},
  volume={555},
  number={7696},
  pages={351--354},
  year={2018},
  publisher={Nature Publishing Group UK London}
}

@article{jovic2022single,
  title={Single-cell {RNA} sequencing technologies and applications: {A} brief overview},
  author={Jovic, Dragomirka and Liang, Xue and Zeng, Hua and Lin, Lin and Xu, Fengping and Luo, Yonglun},
  journal={Clinical and translational medicine},
  volume={12},
  number={3},
  pages={e694},
  year={2022},
  publisher={Wiley Online Library}
}

@article{potter2018single,
  title={Single-cell {RNA} sequencing for the study of development, physiology and disease},
  author={Potter, S Steven},
  journal={Nature Reviews Nephrology},
  volume={14},
  number={8},
  pages={479--492},
  year={2018},
  publisher={Nature Publishing Group UK London}
}

@article{beckman2012impact,
  title={Impact of genetic dynamics and single-cell heterogeneity on development of nonstandard personalized medicine strategies for cancer},
  author={Beckman, Robert A and Schemmann, Gunter S and Yeang, Chen-Hsiang},
  journal={Proceedings of the National Academy of Sciences},
  volume={109},
  number={36},
  pages={14586--14591},
  year={2012},
  publisher={National Academy of Sciences}
}

@article{dutta2022single,
  title={Single-cell profiling of tumour evolution in multiple myeloma—opportunities for precision medicine},
  author={Dutta, Ankit K and Alberge, Jean-Baptiste and Sklavenitis-Pistofidis, Romanos and Lightbody, Elizabeth D and Getz, Gad and Ghobrial, Irene M},
  journal={Nature Reviews Clinical Oncology},
  volume={19},
  number={4},
  pages={223--236},
  year={2022},
  publisher={Nature Publishing Group UK London}
}

@article{proserpio2016single,
  title={Single-cell technologies are revolutionizing the approach to rare cells},
  author={Proserpio, Valentina and L{\"o}nnberg, Tapio},
  journal={Immunology and cell biology},
  volume={94},
  number={3},
  pages={225--229},
  year={2016},
  publisher={Wiley Online Library}
}

@article{chen2014rare,
  title={Rare cell isolation and analysis in microfluidics},
  author={Chen, Yuchao and Li, Peng and Huang, Po-Hsun and Xie, Yuliang and Mai, John D and Wang, Lin and Nguyen, Nam-Trung and Huang, Tony Jun},
  journal={Lab on a Chip},
  volume={14},
  number={4},
  pages={626--645},
  year={2014},
  publisher={Royal Society of Chemistry}
}

@article{du2024single,
  title={Single-molecule scale magnetic resonance spectroscopy using quantum diamond sensors},
  author={Du, Jiangfeng and Shi, Fazhan and Kong, Xi and Jelezko, Fedor and Wrachtrup, J{\"o}rg},
  journal={Reviews of Modern Physics},
  volume={96},
  number={2},
  pages={025001},
  year={2024},
  publisher={APS}
}

@article{han2025solid,
  title={Solid-state spin coherence time approaching the physical limit},
  author={Han, Shuo and Ye, Xiangyu and Zhou, Xu and Liu, Zhaoxin and Guo, Yuhang and Wang, Mengqi and Ji, Wentao and Wang, Ya and Du, Jiangfeng},
  journal={Science Advances},
  volume={11},
  number={9},
  pages={eadr9298},
  year={2025},
  publisher={American Association for the Advancement of Science}
}

@article{hopper2020real,
  title={Real-time charge initialization of diamond nitrogen-vacancy centers for enhanced spin readout},
  author={Hopper, David A and Lauigan, Joseph D and Huang, Tzu-Yung and Bassett, Lee C},
  journal={Physical Review Applied},
  volume={13},
  number={2},
  pages={024016},
  year={2020},
  publisher={APS}
}

@article{xie2021beating,
  title={Beating the standard quantum limit under ambient conditions with solid-state spins},
  author={Xie, Tianyu and Zhao, Zhiyuan and Kong, Xi and Ma, Wenchao and Wang, Mengqi and Ye, Xiangyu and Yu, Pei and Yang, Zhiping and Xu, Shaoyi and Wang, Pengfei and others},
  journal={Science Advances},
  volume={7},
  number={32},
  pages={eabg9204},
  year={2021},
  publisher={American Association for the Advancement of Science}
}

@article{neumann2010single,
  title={Single-shot readout of a single nuclear spin},
  author={Neumann, Philipp and Beck, Johannes and Steiner, Matthias and Rempp, Florian and Fedder, Helmut and Hemmer, Philip R and Wrachtrup, J{\"o}rg and Jelezko, Fedor},
  journal={Science},
  volume={329},
  number={5991},
  pages={542--544},
  year={2010},
  publisher={American Association for the Advancement of Science}
}

@article{bucher2020hyperpolarization,
  title={Hyperpolarization-enhanced {NMR} spectroscopy with femtomole sensitivity using quantum defects in diamond},
  author={Bucher, Dominik B and Glenn, David R and Park, Hongkun and Lukin, Mikhail D and Walsworth, Ronald L},
  journal={Physical Review X},
  volume={10},
  number={2},
  pages={021053},
  year={2020},
  publisher={APS}
}

@article{carver1956experimental,
  title={Experimental verification of the {Overhauser} nuclear polarization effect},
  author={Carver, Thomas R and Slichter, Charles P},
  journal={Physical Review},
  volume={102},
  number={4},
  pages={975},
  year={1956},
  publisher={APS}
}

@article{damadian1971tumor,
  title={Tumor detection by nuclear magnetic resonance},
  author={Damadian, Raymond},
  journal={Science},
  volume={171},
  number={3976},
  pages={1151--1153},
  year={1971},
  publisher={American Association for the Advancement of Science}
}

@article{comcsa2015story,
  title={The story of {MCF-7} breast cancer cell line: 40 years of experience in research},
  author={Com{\c{s}}a, {\c{S}}erban and Cimpean, Anca Maria and Raica, Marius},
  journal={Anticancer research},
  volume={35},
  number={6},
  pages={3147--3154},
  year={2015},
  publisher={International Institute of Anticancer Research}
}

@article{masters2002hela,
  title={{HeLa} cells 50 years on: the good, the bad and the ugly},
  author={Masters, John R},
  journal={Nature Reviews Cancer},
  volume={2},
  number={4},
  pages={315--319},
  year={2002},
  publisher={Nature Publishing Group UK London}
}

@article{koenig1990field,
  title={Field-cycling relaxometry of protein solutions and tissue: implications for {MRI}},
  author={Koenig, Seymour H and Brown III, Rodney D},
  journal={Progress in Nuclear Magnetic Resonance Spectroscopy},
  volume={22},
  number={6},
  pages={487--567},
  year={1990},
  publisher={Elsevier}
}

@article{whitesides2006origins,
  title={The origins and the future of microfluidics},
  author={Whitesides, George M},
  journal={Nature},
  volume={442},
  number={7101},
  pages={368--373},
  year={2006},
  publisher={Nature Publishing Group UK London}
}

@article{squires2005microfluidics,
  title={Microfluidics: Fluid physics at the nanoliter scale},
  author={Squires, Todd M and Quake, Stephen R},
  journal={Reviews of modern physics},
  volume={77},
  number={3},
  pages={977--1026},
  year={2005},
  publisher={APS}
}

@article{allert2022microfluidic,
  title={Microfluidic quantum sensing platform for lab-on-a-chip applications},
  author={Allert, Robin Derek and Bruckmaier, Fleming and Neuling, Nick Ruben and Freire-Moschovitis, Fabian Alexander and Liu, Kristina Song and Schrepel, Claudia and Sch{\"a}tzle, Philip and Knittel, Peter and Hermans, Martin and Bucher, Dominik Benjamin},
  journal={Lab on a Chip},
  volume={22},
  number={24},
  pages={4831--4840},
  year={2022},
  publisher={The Royal Society of Chemistry}
}

@article{henkelman2001magnetization,
  title={Magnetization transfer in {MRI}: a review},
  author={Henkelman, RM and Stanisz, GJ and Graham, SJ},
  journal={NMR in Biomedicine: An International Journal Devoted to the Development and Application of Magnetic Resonance In Vivo},
  volume={14},
  number={2},
  pages={57--64},
  year={2001},
  publisher={Wiley Online Library}
}

@article{nitz1999contrast,
  title={Contrast mechanisms in {MR} imaging},
  author={Nitz, Wolfgang R and Reimer, P},
  journal={European radiology},
  volume={9},
  number={6},
  pages={1032--1046},
  year={1999},
  publisher={Springer}
}

@article{sivelli2025micro,
  title={Micro magnetic resonance spectroscopy for noninvasive metabolic screening of mammalian embryos and oocytes},
  author={Sivelli, Giulia and Barakat, Arthur and Marable, Kathryn B and Gruet, Guillaume and Bitetti, Serena L and Behr, Barry and Lodde, Valentina and Luciano, Alberto Maria and Herrera, Carolina and Blom, Michelle and others},
  journal={Proceedings of the National Academy of Sciences},
  volume={122},
  number={31},
  pages={e2424459122},
  year={2025},
  publisher={National Academy of Sciences}
}

@article{baysoy2023technological,
  title={The technological landscape and applications of single-cell multi-omics},
  author={Baysoy, Alev and Bai, Zhiliang and Satija, Rahul and Fan, Rong},
  journal={Nature Reviews Molecular Cell Biology},
  volume={24},
  number={10},
  pages={695--713},
  year={2023},
  publisher={Nature Publishing Group UK London}
}

@article{mamin2013nanoscale,
  title={Nanoscale nuclear magnetic resonance with a nitrogen-vacancy spin sensor},
  author={Mamin, HJ and Kim, M and Sherwood, MH and Rettner, Charles T and Ohno, K and Awschalom, DD and Rugar, D},
  journal={Science},
  volume={339},
  number={6119},
  pages={557--560},
  year={2013},
  publisher={American Association for the Advancement of Science}
}

@article{staudacher2013nuclear,
  title={Nuclear magnetic resonance spectroscopy on a (5-nanometer)$^3$ sample volume},
  author={Staudacher, Tobias and Shi, Fazhan and Pezzagna, S and Meijer, Jan and Du, Jiangfeng and Meriles, Carlos A and Reinhard, Friedemann and Wrachtrup, Joerg},
  journal={Science},
  volume={339},
  number={6119},
  pages={561--563},
  year={2013},
  publisher={American Association for the Advancement of Science}
}

@article{aslam2017nanoscale,
  title={Nanoscale nuclear magnetic resonance with chemical resolution},
  author={Aslam, Nabeel and Pfender, Matthias and Neumann, Philipp and Reuter, Rolf and Zappe, Andrea and F{\'a}varo de Oliveira, Felipe and Denisenko, Andrej and Sumiya, Hitoshi and Onoda, Shinobu and Isoya, Junichi and others},
  journal={Science},
  volume={357},
  number={6346},
  pages={67--71},
  year={2017},
  publisher={American Association for the Advancement of Science}
}

@article{schmitt2017submillihertz,
  title={Submillihertz magnetic spectroscopy performed with a nanoscale quantum sensor},
  author={Schmitt, Simon and Gefen, Tuvia and St{\"u}rner, Felix M and Unden, Thomas and Wolff, Gerhard and M{\"u}ller, Christoph and Scheuer, Jochen and Naydenov, Boris and Markham, Matthew and Pezzagna, Sebastien and others},
  journal={Science},
  volume={356},
  number={6340},
  pages={832--837},
  year={2017},
  publisher={American Association for the Advancement of Science}
}

\clearpage

\section*{Acknowledgements}
We thank Dominik B. Bucher for his valuable suggestions on the use of hyperpolarization techniques. \textbf{Funding:} This work was supported by the Chinese Academy of Sciences (Grants No.~YSBR-068), the National Natural Science Foundation of China (Grants No.~T2125011, No. 12274396, No. 12404555),  Innovation Program for Quantum Science and Technology (Grants No.~2021ZD0302200, 2021ZD0303204), the China Postdoctoral Science Foundation (Grants No.~2021M703110, No.~2022T150631, No.~2023M743399, No.~2024M753084), the Postdoctoral Fellowship Program of CPSF (Grants No.~GZB20240717, No.~GZB20240718), and the Fundamental Research Funds for the Central Universities, New Cornerstone Science Foundation through the XPLORER PRIZE.
\textbf{Author contributions:} J.D. and F.S. supervised the project and proposed the idea. Z.Z., F.S. and J.D. designed the experiments. Z.Z., Q.S., S.X. and T.X. prepared the setup. X.Y. and Y.W. prepared the diamond sample. Q.S. and J.S. prepared the cell samples. Z.Z., Q.S. and S.X. performed the experiment and the simulation. Q.S. and Z.Z. performed the primary data acquisition. Z.Z., Q.S., Q.H., F.S. and J.D. wrote the manuscript. All authors analysed the data, discussed the results and commented on the manuscript.
\textbf{Competing interests:} The authors declare that they have no competing interests. 
\textbf{Data and materials availability: } All data needed to evaluate the conclusions in the paper are present in the paper and/or the Supplementary Materials.

\pagebreak

\end{document}